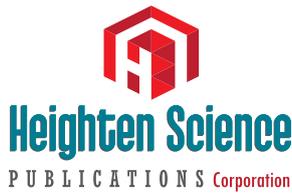

**PUBLICATIONS Corporation**

**Case Report**

# A Short Synthesis Concerning Biological Effects and Equivalent Doses in Radiotherapy

## Cyril Voyant[1,2]* and Daniel Julian[3]

[1]University of Corsica, CNRS UMR SPE 6134, (Campus Grimaldi, 20250 Corte), France
[2]Hospital of Castelluccio, Radiotherapy Unit, BP 85, 20177 Ajaccio, France
[3]CGS cancérologie du grand Montpellier, 34000 Montpellier, France

**\*Address for Correspondence:** Dr. Cyril Voyant, University of Corsica, CNRS UMR SPE 6134, (Campus Grimaldi, 20250 Corte), France, Tel: +33 495293666; Fax: +33 495293797; Email: voyant@univ-corse.fr

## ABSTRACT

The limits of classical equivalent computation based on time, dose, and fractionation (TDF) and linear quadratic models have been known for a long time. Medical physicists and physicians are required to provide fast and reliable interpretations regarding the delivered doses or any future prescriptions relating to treatment changes. In this letter, we propose an outline related to the different models usable for equivalent and biological doses that are likely to be the most appropriate. The used methodology is based on: the linear-quadratic-linear model of Astrahan, the repopulation effects of Dale, and the prediction of multi-fractionated treatments of Thames.

## PROBLEMS OF THE BIOLOGICALLY EQUIVALENT DOSE

It has long been known that radiation biology plays an important role and it is necessary for radiotherapy treatments. The radiation effects on normal and malignant tissues after exposure range from a femtosecond to months and years thereafter [1,2]. Therefore, to optimize treatment, it is crucial to explain and understand these mechanisms [3-5]. Providing a conceptual basis for radiotherapy and identifying the mechanisms and processes that underlie the tumor and normal tissue responses to irradiation can help to explain the observed phenomena [6]. Examples include understanding hypoxia, reoxygenation, tumor cell repopulation, or the mechanisms of repair of DNA damage [3,7,8]. The different biological effects of radiation should be divided into several phases: physical (interaction between charged particles and tissue atoms), chemical (the period during which the damaged atoms and molecules react with other cellular components in rapid chemical reactions), and biological (impact of the generated lesions on the biological tissue [4]). The following section describes the models most often used in radiotherapy. These are simplistic models that actual treatments are based, and that is validated and approved [9-12].

## REFERENCE MODELS

Numerous models exist to evaluate the biological equivalent dose, but the two most common are the nominal standard dose (NSD [13]) and linear quadratic (LQ [9]) models. The NSD uses the power law described in equation 1 below ($D_{tot}$ is the tolerance dose of the tissue, $NSD$ is a constant, $n$ and $t \in \mathbb{R}^+$, $N$ the number of fractions, and $T$ the overall treatment time). However, this model has been often criticized [14]. In short, some researchers consider and have even shown that the NSD formula is not a valid description for all tumors and normal tissues; instead, they maintain that the model incorrectly describes the effects of fraction number and treatment duration.

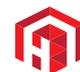

$$\in \mathbb{R}^+ \qquad D_{tol} = NSD.N^n.T^t \tag{1}$$

The LQ model is most frequently used in the radiotherapy units. It allows the equivalent dose to be easily evaluated for different fractionations. This concept involves the $\frac{\alpha}{\beta}$ ratio, as shown in equation 2 below ($D$ is the total dose for a fraction size of $d$ gray).

$$EQD_2 = D.\frac{d + \frac{\alpha}{\beta}}{2 + \frac{\alpha}{\beta}} \qquad\qquad EQD_2 = D.\frac{d + \frac{\alpha}{\beta}}{2 + \frac{\alpha}{\beta}} \tag{2}$$

$EQD_2$ Is the dose obtained using a 2Gy fraction dose, which is biologically equivalent to the total dose $D$ given with a fraction dose of $d$ gray the values of $EQD_2$ may be added in separate parts in the treatment plan. This formula may be adapted to fraction doses other than 2Gy.

### Limitations of the LQ model

The LQ model is frequently used for modeling the effects of radiotherapy at low and medium doses per fraction for which it appears to fit clinical data reasonably well. The main disadvantage of the LQ approach is that the overall time factor is not taken into account, because in radiotherapy it is regarded to be more complex than previously supposed [3]. It is indeed very difficult to include this *parameter* in the LQ equation. However, a technique may be used to integrate a penalty term in Equation 2. Thus, for $T_{stop}$ days off treatment, the dose recovered would be $T_{stop}.D_{prol}$, where $D_{prol}$ is the proliferation factor (in Gy/day; for example, 0.22 for laryngeal edema or 0.15 for rectosigmoid *complications*). This methodology is essentially validated for discontinuation during treatment. As a general rule, the main limitations of using the LQ model are linked to repopulation (LQ *doesn't* take into account the dose protraction), bi-fractionated treatments and high-dose fractions (continuously bending survival curve versus linear behavior observed at least in some cell lines). Other more sophisticated models, however, take into account these weaknesses. We will later see that *the* LQ model requires further theoretical investigation, especially in terms of the biologically effective dose (BED). The next section describes the theoretical methodology that we propose to compute the BED.

## THE EXISTING MODELS

The BED (introduced by Fowler [9]) is a mathematical concept used to illustrate the biological effects observed after irradiation. In addition to being easily computable (BED= physical dose x relative efficiency), this notion is interesting because two irradiations with the same BED generate the same radiobiological effects. For this reason, it is easy to compare treatments with different doses, fractionations, and overall times.

### Target volume models

Let us examine two different treatment cases separately. The first involves treatments with a high-dose fraction (one treatment per day, the fraction size $d$ is greater than the $d_t$ limit; [15]), which requires a linear quadratic linear (LQL) model. The second case relates to other treatments ($d < d_t$), where the LQ model is applicable to daily multi-fractionation [16].

#### a. The $d > d_t$ case

When the dose per fraction ($d$) is greater than the LQL threshold ($d_t \sim 2 \ \alpha/\beta$), the BED is computed using Equation 3 below (one fraction permitted per day). This template regroups Astrahan's high-dose model [17] and Dale's repopulation model [18] ($n$ is the number of fraction, $\theta(x)$ the Heaviside function, $\frac{\gamma}{a}$ the parameter of the LQL model and $T_{pot}$ the potential doubling time in day).

$$BED = n.\left(d_t.\left(1+\frac{d_t}{\alpha/\beta}\right)+\frac{\gamma}{a}.(d-d_t)\right)-\theta(T-T_k).\frac{\ln(2)}{\alpha.T_{pot}}.(T-T_k) \tag{3}$$

The second term used in this equation is useful only when the overall time $T$ is greater than the $T_k$ value (kick-off time). If this threshold is not achieved, the tumor cells are considered to be non-proliferative (early hypoxia).

### b.   The $d \leq d_t$ case

When the fraction dose is low, it is recommended to use the standard BED equations while considering one or more fractions per day (Equation 4). This methodology follows the model of Thames [19], who introduces the repair factor $H_m$ related to the amount of unrepaired damage (Equation 5). If the inter-fraction interval is reduced below the full repair interval (between 6 hours and 1 day), the overall damage from the whole treatment is increased because the repair of damage due to one radiation dose may not be complete before the next fraction is given. $H_m$ is LQ model correction taking account the poly-fractionation, $m$ the number of fraction per day, $\phi$ the incomplete repair and $\mu$ the parameter adjustment necessary to take into account the poly-fractionation in the model LQ in hours$^{-1}$.

$$BED = n.d.\left(1+(1+H_m).\frac{d}{\alpha/\beta}\right)-\theta(T-T_k).\frac{\ln(2)}{\alpha.T_{pot}}.(T-T_k) \tag{4}$$

$$H_m = \left(\frac{2}{m}\right).\left(\frac{\phi}{1-\phi}\right).\left(m-\frac{1-\phi^m}{1-\phi}\right) \text{ and } \phi = e^{(-\mu.\Delta T)} \tag{5}$$

It is worth noting that in the case of mono-fractionation, the $H_m$ factor is null. These equations only relate to the target volume calculation. For the organs at risk, the kick-off time is not relevant, meaning that it is necessary to use a repopulation-specific approach.

## Models for organs at risk

As in the precedent section on target volumes, this section similarly separates high and low doses per fraction. The BED formulae are almost equivalent to the target volume model; only the terms relating to the lack of dose by proliferation are modified.

### a.   The $d > d_t$ case

To understand this methodology, it is necessary to consult Van Dyk's law [20]. The kick-off time is no longer considered, with the recovered dose $\left(D_{rec} = \frac{\ln(2)}{\alpha.T_{pot}}\text{ in Gy / day}\right)$ instead being added. The global model is described in Equation 6 below.

$$BED = n.\left(d_t.\left(1+\frac{d_t}{\alpha/\beta}\right)+\frac{\gamma}{a}.(d-d_t)\right)-D_{rec}.T \tag{6}$$

### b.   The $d \leq d_t$ case

In the case of low doses per fraction, the methodology is similar to the target volume model: the $H_m$ parameter (Equation 5) is nonetheless required, which allows us to take into account more than one fraction per day. As seen in the Equation 7 below, the recovered dose is used as in the previous case.

$$BED = n.d.\left(1 + (1+H_m).\frac{d}{\alpha/\beta}\right) - D_{rec}.T \tag{7}$$

## Proposztion for computational methods for the equivalent dose

The standard models used for the equivalent dose as based on the LQ approach are easily exploitable. The main formulation of the model (Equation 2) can be obtained by considering the general formula described in the Equation 8 as follows.

$$D_1 = D_2.\frac{(\alpha/\beta + d_2)}{(\alpha/\beta + d_1)} \tag{8}$$

This equation may be validated using BED methodology. Considering the BED of two treatments to be equal, it appears that a simple relation links the two overall doses $D_1$ ($=n_1.d_1$) and $D_2$ (= $n_2.d_2$). The detail of this procedure is shown in the Equation 9 below.

$$BED_1 = n_1.d_1.\left(1 + .\frac{d_1}{\alpha/\beta}\right) = BED_2 = n_2.d_2.\left(1 + \frac{d_2}{\alpha/\beta}\right) \tag{9}$$

In the case of more sophisticated BED formulations, it is not easy to determine a simple formula linking the $D_1$ and $D_2$ doses, as recovery and repopulation significantly complicate the computational principle. Most of the existing software that uses the overall time correction does not calculate the equivalent dose; instead, it only provides the BED for the chosen treatments. In clinical use, it is more valuable for the physician or physicist to work with the equivalent dose in standard fractionation. In this context, the proposed methodology is based on an innovative algorithm, which allows a cost function extremum to be determined based on BED modeling. To explain this methodology, it is necessary to consider two irradiations (Indices 1 and 2), which are defined by a fraction number ($n$), dose per fraction ($d$), and days of discontinuation ($ja$). The corresponding BED is noted as BED1 ($n_1,d_1,ja_1$) and BED2 ($n_2,d_2,ja_2$), while the cost function $f$ is defined in Equation 10 as follows.

$$f\left(n_1, d_1, ja_1, n_2, d_2, ja_2\right) = \left| BED_1(n_1, d_1, ja_1) - BED_2(n_2, d_2, ja_2) \right| \tag{10}$$

In clinical use, it is desirable to compare a radiotherapy trial with one that is performed in a conventional manner (generally with 2 Gy per fraction without interruption). This concept of a reference dose simplifies the issue, as it is thus possible to dispense with the days off treatment and multi-fractionation per day in relation to the reference treatment. The following example concerns a tumor case with a dose per fraction less than $d_t$ (second part of the target volume model), while the cost function, $f$, is given in Equation 11. Concerning the three other cases examined in previous sections, a similar relationship is found.

$$f\left(n_{ref}, d_{ref}, n, d, ja\right) = \left| \begin{array}{l} n_{ref}.d_{ref}.\left(1 + \frac{d_{ref}}{\alpha/\beta}\right) - \theta\left(T_{ref} - T_k\right).\frac{\ln(2)}{\alpha.T_{pot}}. \\ (T_{ref} - T_k) - n.d.\left(1 + (1+H_m).\frac{d}{\alpha/\beta}\right) - \theta(T - T_k).\frac{\ln(2)}{\alpha.T_{pot}}.(T - T_k) \end{array} \right| \tag{11}$$

The global treatment duration can be seen to be directly associated with the fraction number and days off during radiotherapy. Following Equation 11, the 2Gy-per-fraction equivalent dose (EQD$_2$) for standard treatment with the characteristics $n, d, ja$ is given by the algorithm shown in Equation 12.

$$\begin{cases} \underset{n_{ref} \in \mathbb{R}^+}{\arg\min} \, f\left(n_{ref}, 2, n, d, ja\right) = n_0 \\ EQD_2 = 2.n_0 \end{cases} \tag{12}$$

All of the results obtained in this section were implemented using a Matlab® standalone application known as LQL-equiv.

## SYNTHESIS OF THE DOSE EQUIVALENT MODELS

The LQL_Equiv software was developed in collaboration by the CHD Castelluccio radiotherapy unit in Ajaccio and the University of Corsica. It is a free software released under the GNU license [21]. The source codes, executable file, help files, and license terms are available at http://cyril-voyant.univ-corse.fr/LQL-Equiv_a34.html. Before installing this software, it is advisable to refer to the installation guide and to download and execute Matlab Component Runtime (MCR 32 bits, version 7.15 or later). This latter step is necessary since the application was programmed using the GUI Matlab® software (32 bits, v. 7.12) and deployed with the Matlab Compiler® (v. 4.12), which use MCR (a standalone set of shared libraries enabling the execution of Matlab® applications on a computer without an installed version of Matlab®). Users of the LQL_Equiv software are advised to provide us with comments on the software, its libraries (biological parameters for each organ or tumor type), or any bugs so as to allow us to develop the software. It should be noted that the application requires Microsoft Windows® (the resolution and colors are for Vista or later versions). It is important to note that all of the parameters used for calculating the equivalence are available on the graphical interface. Using Matlab™ and the downloadable source codes, it is easy to modify or complete these parameters. It is also possible to contact the software authors to assist in developing the software. LQL_Equiv is in direct competition with TDF Plan developed by Eye Physics LLC, which proposes a multitude of parameters. However, the software is dedicated to the calculation of BED and is not really consistent with the reference equivalent dose. Moreover, we aimed to develop ergonomic software with minimum of adjustable parameters, which ultimately complicate the interpretation of the output. These two approaches are nevertheless complementary; for more information about the different models used, refer to the TDF Plan website (http://www.eyephysics.com/tdf/Index.htm). Table 1 presents a comparison between outputs of the standard calculation models described in section II (LQ without proliferation and $\alpha/\beta = 10$ for oral mucosa and 2 for others) and the LQL-equiv software. The difference between the two approaches is substantial. The overall time effect and unusual doses per fraction result in completely different outputs. The maximum difference is close to 25%; this value is linked to the cell repopulation of prostate cancer. In this case, the non-specific methods are certainly not usable.

## CONCLUSION

In this letter, we have exposed the compiling results of various published LQ model modifications, which have been modified to be better suited for specialized radiotherapy techniques such as hypo- or hyperfractionation. The LQ model was modified to take into account multi-fractionation, repopulation, high-dose fractions, and overall time. Moreover, we propose a software program (LQL_equiv), integrating all of these concepts regarding the main organs at risk or target volumes. Finally, this worklow permits the obtained results to be compared and validated against other "homemade" models, with the purpose of harmonizing practices in interested centers. However, it is essential not to consider models as "general biological rules", parameters and outputs uncertainties can be very large; this phenomenon is related to the number of regression parameters (parsimony principle) and to the data snooping (e.g. failure to adjust existing statistical models when applying them to new datasets).

**Table 1:** Comparison between the outputs of the LQL_Equiv and standard calculation models (LQ without proliferation and with $\alpha/\beta$ =10 for oral mucosa and 2 for others). Bold font is used to show differences >5%.

| Treatements | | Organs at risk | Target volums |
|---|---|---|---|
| 10x3Gy | | Spinal cord | Prostate (metastasis) |
| | classical output (Gy) | 37.5 | 37.5 |
| | LQL-equiv output (Gy) | 37.5 | 36 |
| | Difference (Gy / %) | -0 / -0% | -1.5 / -4% |
| 10x3Gy | | Spinal cord | Breast (metastasis) |
| | classical output (Gy) | 37.5 | 37.5 |
| | LQL-equiv output (Gy) | 37.5 | 38.2 |
| | Difference (Gy / %) | -0 / -0% | 0.7 / 1.9% |
| 1x8Gy | | Spinal cord | Prostate (metastasis) |
| | classical output (Gy) | 20 | 20 |
| | LQL-equiv output (Gy) | 16 | 16.8 |
| | Difference (Gy / %) | -4 / **-20%** | -3.2 / **-16%** |
| 10x3Gy | | Brain | Breast (metastasis) |
| | classical output (Gy) | 37.5 | 37.5 |
| | LQL-equiv output (Gy) | 43.5 | 38.2 |
| | Difference (Gy / %) | 6 / **16%** | 0.7 / 1.9% |
| 1x8Gy (1 month gap time) 1x8Gy | | Spinal cord | Prostate (metastasis) |
| | classical output (Gy)) | 40 | 40 |
| | LQL-equiv output (Gy) | 32 | 33.3 |
| | difference (Gy / %) | -8 / -4.63% | -6.7 / **16.7%** |
| 5x4Gy | | Pericardium | Lung (metastasis) |
| | classical output (Gy) | 30 | 30 |
| | LQL-equiv output (Gy) | 37.5 | 23.3 |
| | Difference (Gy / %) | 7.5 / **25%** | -6.7 / **-22.3%** |
| 20x2Gy (1 week gap time) 10x2Gy | | Oral mucosa | Oropharynx |
| | classical output (Gy) | 60 | 60 |
| | LQL-equiv output (Gy) | 54.4 | 53 |
| | difference (Gy / %) | -5.6 / **-9.3%** | -7 / **-11.7%** |
| 22x1.8Gy (bi-fractionated) | | Oral mucosa | Oropharynx |
| | classical output (Gy) | 38.9 | 38.9 |
| | LQL-equiv output (Gy) | 45 | 36 |
| | difference (Gy / %) | 6.1 / **15.7%** | -2.9 /**-7.4%** |
| 25x1.8Gy then 15x2Gy | | Rectum | Prostate |
| | classical output (Gy) | 72.7 | 72.7 |
| | LQL-equiv output (Gy) | 71 | 73 |
| | Difference (Gy / %) | -1.7 / -2.3% | 0.3 / 0.4% |
| 20x2.5Gy (4 fraction/week) | | Lung | Breast |
| | classical output (Gy) | 56.2 | 56.2 |
| | LQL-equiv output (Gy) | 62.9 | 56.8 |
| | difference (Gy / %) | 6.7 / **11.9%** | 0.2 / 0.3% |
| 4x4.5Gy (2 week gap time) 4x4Gy | | Optic chiasma | Glioblastoma |
| | classical output (Gy) | 53.2 | 53.2 |
| | LQL-equiv output (Gy) | 42.8 | 47.4 |
| | difference (Gy / %) | -10.4 / **-19.5%** | -5.8 / **-10.9%** |
| 28x1.8Gy (1 week gap time) | | Skin (early) | Breast |
| | classical output (Gy) | 47.9 | 47.9 |
| | LQL-equiv output (Gy) | 47.6 | 42.3 |
| | difference (Gy / %) | -0.3 / 0.6% | -5.6 / **-11.7%** |